\documentstyle[aps,prl,twocolumn,graphicx]{revtex}

\begin{document}

\setlength{\unitlength}{1cm}

\title{Unified Quantum Mechanical Picture for Confined Spinons\\
 in Dimerized and Frustrated Spin $S=1/2$ Chains}

\author{G\"otz S.~Uhrig, Friedhelm Sch\"onfeld}
\address{Institut f\"ur Theoretische Physik, Universit\"at zu K\"oln,
 Z\"ulpicher Stra\ss e 77, 50937 K\"oln, Germany}

\author{Markus Laukamp, Elbio Dagotto}
\address{National High Magnetic Field Laboratory and Department of
Physics,\\
 Florida State University, Tallahassee, Florida 32306, USA}

\draft

\maketitle
\begin{abstract}
A quantum mechanical picture is presented to describe the
behavior of confined spinons in a variety of $S=1/2$ chains. 
The  confinement is due to dimerization and frustration and it
manifests itself
as a nonlinear potential $V(x)\propto |x|^b$, centered at 
chain ends ($b\le 1$) or produced by modulation kinks ($b> 1$).
 The calculation extends to weak or zero frustration some previous
 ideas valid for spinons in strongly frustrated spin chains.
The local magnetization patterns
of the confined spinons are calculated. A (minimum) enhancement of the
local moments of about 11/3 over a single $S=1/2$ is found.
Estimates for excitation energies and binding lengths are obtained.
\end{abstract}

\section{Introduction}
Dimerized and frustrated spin chains have attracted considerable interest
in recent years. This is due to the recent emergence of a variety of
experimental quasi-one-dimensional systems containing localized electrons that 
can be described by spin chains. Among these systems there is a sizable
number which are gapful due to dimerization in their low temperature
phase, such as CuGeO$_3$ \cite{hase93a},
$\alpha'$-NaV$_2$O$_5$  \cite{isobe96}, (VO)$_2$P$_2$O$_7$ \cite{garre97a},
Cu(NO$_3$)$_2\cdot2.5$H$_2$O \cite{bonne83},
CuWO$_4$ \cite{lake96},
 and Cu$_2$(C$_2$H$_{12}$N$_2$)$_2$Cl$_4$ \cite{chabo97}.
The first two dimerize as a consequence
of the coupling to the lattice degrees of freedom, whereas the other compounds are intrinsically dimerized, i.e. the dimerization does not depend on
temperature. The effect of doping on these substances
represents an interesting issue, in particular since it pertains also
to defects such as missing spins, broken chains and chains of
finite length. 

As a general rule, defects in low-dimensional antiferromagnetic
spin systems which can be described by a RVB-type ground state \cite{liang88}
involve $S=1/2$ states in their vicinity \cite{marti97}. 
The appearance of a certain impurity spin at the edges of spin chains
can also be discussed in the framework of a nonlinear $\sigma$ model 
for general spin $S$ \cite{ng94}. We restrict ourselves to $S=1/2$,
weakly dimerized spin chains. The basic idea there is that
without the defect each spin has a partner with which a singlet
is formed. If the defect removes one of the partners, the other becomes
a free $S=1/2$ spin which we will henceforth call spinon to
distinguish it from the other singlet forming spins. We understand that
this spinon comprises also some dressing of the bare $S=1/2$ spin, i.e.
it is not localized at just one site.
Since the singlet pairing is not static in the
RVB-picture the spinon is able to move some distance away from the
defect. In a critical, gapless system,
it will be delocalized. This delocalization, however, disappears
 as soon as the couplings are modulated.

It is the purpose of this work to elucidate in which
way an explicit dimerization acts as a confining potential for the 
spinon motion leading to its localization.
The generic Hamiltonian reads
\begin{equation}
H = J\sum_i\left( (1+(-\delta)^i) \vec{S}_i \vec{S}_{i+1} +
\alpha \vec{S}_i \vec{S}_{i+2}\right)
\end{equation}
where $\delta$ parametrizes the dimerization and $\alpha$
the relative frustration by next-nearest neighbor coupling.
Previous works, e.g. \cite{khoms96,affle97,marti96b,els98},
viewed dimerization already as confining potential.
We understand that there is no confinement without dimerization ($\delta=0$).
Here we will develop a quantum mechanical picture that treats the cases of
small and large frustration on equal footing. Although this 
picture is not exact in all details, nevertheless it is able to reproduce the
main features of the problem on a semi-quantitative level.
 The physical quantities 
considered here are binding energies and local magnetizations.
In addition, it will be shown that the confinement
depends on the degree of frustration and it will be sublinear
 in the region of low frustration.

\begin{figure}
\begin{picture}(8.2,5.2)
\put(0,0.5){\includegraphics[width=8cm]{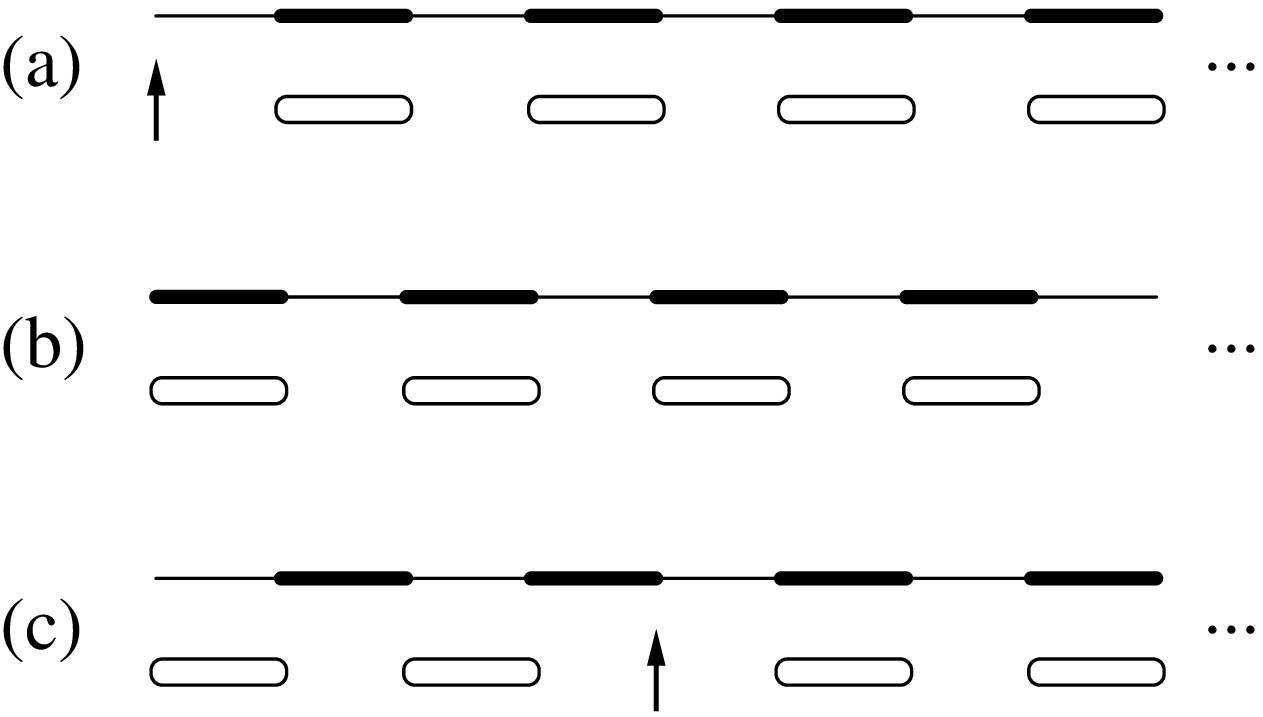}}
\end{picture}
\caption[]{
Three possible configurations at chain ends.
The thick (thin) solid lines stand for strong (weak) bonds; the open
eyelets stand for singlets, the arrow for an unpaired spin.
(a) weak bond at the chain end corresponds to a free spin; (b) strong
bond at the chain end corresponds to no free spin; (c) configuration
after two hops of the free spin from (a). Note the misaligned first
two singlets at the strong bonds.}
\label{fig-idea}
\end{figure}

For the sake of concreteness, the case of a chain end is
illustrated in Fig.\ref{fig-idea}.
Dimerization localizes spin singlets mostly at the strong bonds.
If in a dimerized chain one spin is missing its singlet partner is 
freed. If the missing spin had a weak bond to the left (assuming without loss
of generality horizontal chains) and a strong
bond to the right, the free spin situation corresponds to the one in
Fig.\ref{fig-idea}(a). The configuration in Fig.\ref{fig-idea}(b)
is then the reflected configuration found on the left of the missing
spin. If the missing spin had a strong bond to the left and a weak
bond to the right, the free spin situation is found to the left
of the missing spin. The configuration in Fig.\ref{fig-idea}(a)
is then a reflected image whereas the configuration in Fig.\ref{fig-idea}(b)
is found to the right of the missing spin.
Fig.\ref{fig-idea}(c) illustrates the situation after two hops of the
free spin away from its origin in (a). The crucial point is that the
singlets to the left of the spin in Fig.\ref{fig-idea}(c) are no longer
at the strong bonds. This implies an energy loss which can be viewed
as an attractive potential which ties the free spin to its origin.
Besides chain ends also solitonic modulations, such as
kink defects, will  be considered in this paper. They can also
be viewed as spinon traps with confining potentials.

The article is organized in the following way. In the next section, 
the short-range RVB spinon states that determine the main part of the 
correlations will be introduced. Subsequently, the forms of the kinetic
 and the potential energies which govern the spinon dynamics will be 
discussed, i.e. the Schr\"odinger equation of the problem will be setup. 
 In the fourth  and the fifth sections the quantitative results will be
discussed and  compared against computational calculations  for chain
ends and for kink-modulated chains, respectively. 
A summary will conclude the article.

\section{States: Norm and Magnetization}
Let us denote by $|i\rangle$ an up-spinon at site  $2i+1$; the other spins
are all paired to nearest-neigbor singlets. In this convention, the state in
Fig.\ref{fig-idea}(a) is denoted by $|0\rangle$ and the state in
Fig.\ref{fig-idea}(c) is denoted by $|2\rangle$. These states are not 
orthogonal but their overlap \cite{caspe84,liang88,mulle98} is given by 
\begin{equation}
\label{overlap}
\langle i|j\rangle = \left(-\frac{1}{2}\right)^{|i-j|}\ .
\end{equation}
This overlap arises from a N\'eel type sequence of up-spins and
down-spins between the two spinons. 
Let us consider a state 
\begin{equation}
|v\rangle =  \sum_{i=0}^\infty a_i (-1)^i |i\rangle
\end{equation}
for which we aim to obtain a continuum description. This means that we assume
that for the low energy behavior it is sufficient to treat $a_i$
as a slowly varying function of $i$. 
Note that we introduced the factor $(-1)^i$ to focus on the
energetically low-lying states. This can be seen most easily
for $\alpha=1/2$ \cite{shast81,caspe84,mulle98}.
In order to define a normalized
wave function $\psi(x)$ let us calculate the norm of $|v\rangle$
\begin{eqnarray}
\langle v|v\rangle & =&  \sum_{i,j} a_i^* a_j
\left(\frac{1}{2}\right)^{|i-j|}
\nonumber \\
&\approx& \sum_i |a_i|^2 \left(-1 + 2 \sum_{j=0}^\infty 2^{-j} \right)
\nonumber \\
\label{norm}
& = & 3 \sum_i |a_i|^2 \ .
\end{eqnarray}
The approximate step is valid if $a_i$ varies in fact slowly with $i$.
Eq.(\ref{norm}) tells us that passing to a normalized continuous 
wave function $\psi(x)$ means $i \to x/2$ and $a_i \to
\sqrt{2/3}\psi(x)$.
The factor 2 introduced here enables the replacement of $x$ by the
 actual site numbers at the end of the calculation.

Now we calculate the local magnetizations $m_l := \langle S^z_l \rangle$.
\begin{equation}
\langle v| S^z_l|v\rangle = \frac{(-1)^{l+1}}{2} \!\!\!\!\!\!
\sum_{\small
\begin{array}{l}
2i+1\le l \le 2j+1 \ \mbox{or}\\
2j+1\le l \le 2i+1
\end{array}
}\!\!\!\!\!\!
 a_i^* a_j
 \left(\frac{1}{2}\right)^{|i-j|}  \ .
\end{equation}
A contribution is found only if $S^z_l$ is situated {\em between}
the two spinons which equals one half of the overlap (\ref{overlap}).
For $l$ odd one finds
\begin{eqnarray}
\langle v| S^z_l|v\rangle 
&\approx& |a_{(l-1)/2}|^2 \frac{1}{2} \left(-1 + 2\sum_{n,m =0}^\infty 
\left(\frac{1}{2}\right)^{n+m}\right)
\nonumber \\
&=& \frac{7}{2} |a_{(l-1)/2}|^2
\nonumber \\
&=& \frac{7}{3} |\psi(l)|^2 
\label{resodd1}
\end{eqnarray}
where $n$ and $m$ stand for half the distance to site $l$ on the left
and on the right.
We do not take the chain end into account, i.e. we 
ignore the fact that summation for small $l$ is truncated on the left.
For even sites $l$ one obtains
\begin{eqnarray}
\langle v| S^z_l|v\rangle 
&\approx& - \frac{1}{2} |a_{l/2}|^2 2\frac{1}{2} \sum_{n,m =0}^\infty
2^{-n-m}
\nonumber \\
&\approx& - 2 |a_{l/2}|^2
\nonumber \\
\label{reseven1}
&=& - \frac{4}{3} |\psi(l)|^2 \ .
\end{eqnarray}
With the above calculation the local magnetizations have been linked
to the probability of finding the spinon at a given site $l$.
The sum $\sum_l m_l$ equals to one half as it has to be for a global
$S=1/2$ state. Note that to obtain this result one has to sum even and odd
sites separately, and that $l$ changes, thus, by two from site to site.

Summing the moduli $\sum_l |m_l|$ one gets immediately 
$\frac{11}{3}\frac{1}{2}$. This means that the antiferromagnetic
correlations induce total local moments that correspond to the
moments of $11/3=3.667$ independent spins $S=1/2$. This enhancement factor
of $11/3$ illustrates why even a low concentration of dopants
may induce considerable antiferromagnetism in a compound
\cite{fukuy96,marti97,lauka98}.
Moreover, this enhancement is not due to criticality but it is already
built-in in the short-range RVB state.
The possible coexistence of dimerization and alternating local
magnetization was nicely demonstrated by Fukuyama {\it et al.}
by analysis of the corresponding phase hamiltonian \cite{fukuy96}.
Impurity-induced antiferromagnetism is already extensively investigated
for spin ladders \cite{motom96,fukuy96b,sigri96,iino96,mikes97,imada97}.

In the next paragraphs we will refine the calculation of the norm and 
the local magnetization by considering indeed a discrete wave function.
We would like to remove the non-orthogonality of our basis states
in order to be on safer ground for the subsequent reasoning. 
Moreover, certain boundary effects can be captured by this procedure.

By  Gram-Schmidt orthogonalization we find the orthonormal basis
\begin{mathletters}
\begin{eqnarray}
|w_0\rangle &:=& |0\rangle
\\
|w_{i>0}\rangle &:=&(-1)^i \frac{2}{\sqrt{3}} (|i\rangle + |i-1\rangle)\ .
\end{eqnarray}
\end{mathletters}
The above definitions simplify any norm calculation. The magnetization
calculation becomes a bit more tedious. For odd $l=2j+1$ we have
\begin{mathletters}
\label{lodd}
\begin{eqnarray}
\langle w_{j+1}| S^z_{2j+1} |w_{j+1} \rangle &=& -1/6 \\
\langle w_{j}| S^z_{2j+1} |w_{j} \rangle &=& 1/2 \quad \mbox{for} \ j=0 \\
\langle w_{j}| S^z_{2j+1} |w_{j} \rangle &=& 1/3 \quad \mbox{for} \ j>0 \\
\langle w_{j}| S^z_{2j+1} |w_{i} \rangle &=& 2^{-1-j+i} \quad \mbox{for} \
j>i>0 \\
\langle w_{j}| S^z_{2j+1} |w_0 \rangle &=& 2^{-j}/\sqrt{3}
\end{eqnarray}
\end{mathletters}
and the analogous formulae hold for $i \leftrightarrow j$.
All other expectation values are zero.
From the relations (\ref{lodd}) one finds for a normalized
state $|v\rangle = \sum_{i=0}^\infty b_i |w_i\rangle$
\begin{mathletters}
\label{resodd2}
\begin{eqnarray}
\langle v| S^z_1 | v\rangle &=& \frac{1}{2} |b_0|^2-\frac{1}{6} |b_1|^2
\\
\langle v| S^z_{2j+1} | v\rangle &=& 
\frac{1}{3} |b_j|^2-\frac{1}{6} |b_{j+1}|^2+\nonumber \\
&+& \mbox{Re}\ b^*_j 2^{-j} \left( \frac{2}{\sqrt{3}} b_0 + \sum_{i=1}^{j-1}
b_i 2^i \right) \ .
\end{eqnarray}
\end{mathletters}
Repeating the same for even $l=2j$ we arrive at
\begin{mathletters}
\label{leven}
\begin{eqnarray}
\langle w_{j}| S^z_{2j} |w_{j} \rangle &=& 1/3 \\
\langle w_{j}| S^z_{2j} |w_{i} \rangle &=& - 2^{-1-j+i} 
\ \mbox{for} \ 0 < i < j \\
\langle w_{j}| S^z_{2j} |w_0 \rangle &=& - 2^{-j}/\sqrt{3} \ .
\end{eqnarray}
\end{mathletters}
From  Eqs. (\ref{leven}) we obtain
\begin{equation}
\label{reseven2}
\langle v| S^z_{2j} | v\rangle = \frac{1}{3} |b_j|^2-
\mbox{Re}\ b_j^* 2^{-j} \left(\frac{2}{\sqrt{3}}b_0 + \sum_{i=1}^{j-1}
b_i 2^i  \right) \ .
\end{equation}

Note that for $l\gg0$ and slowly varying $b_i$ one re-obtains, of course,
the results Eqs.(\ref{resodd1}) and (\ref{reseven1}). This concludes
the calculation of the local magnetizations.

\section{Kinetic and Potential Energy}
In this section we address the important issue of how the simple spinons
introduced in the previous paragraph move and how they are attracted by
the chain end. We will distinguish two regimes of frustration (i) 
$0.5 \ge \alpha > \alpha_c=0.241$ \cite{julli83,okamo92} and (ii)
$\alpha \le \alpha_c$. 

In regime (i) we know that the spinons
have a finite mass and that they have a quadratic minimum in their dispersion
\cite{shast81,caspe84,mulle98}. Moreover, the strongly frustrated chains
display spontaneous symmetry breaking of the translational symmetry, i.e.
spontaneous dimerization occurs. The actual value of the gap (and the
dimerization), however, is very small 
up to $\alpha=0.35$ 
\cite{chitr95,yokoy97} (see also Fig.\ref{fig-gap}).

In regime (ii) the spinon dispersion is linear in the wave vector
$k$ \cite{cloiz62}. Hence we assume $\omega(k)=v_S |k|$ where 
$v_S$ is the spin wave velocity which is roughly given by
\begin{equation}
\label{spinwave}
v_S = \frac{\pi}{2}(1-1.12\alpha)
\end{equation}
according to Ref.\cite{cloiz62,fledd97}.
The multiplication by $|k|$ in reciprocal space  can be visualized
as the consecutive multiplications $\mbox{sgn}(k) \cdot k$. To multiply
by $k$ corresponds to $-i\frac{\partial}{\partial x}$ in real space,
while a multiplication by $\mbox{sgn}(k)$ corresponds to the convolution with
the principal value of $i/(\pi x)$. Then,  $|k|$ in
reciprocal space corresponds to the  convolution 
\begin{mathletters}
\begin{eqnarray}
H_{\rm kin} \psi(x) &=& -i\frac{\partial}{\partial x}
{\cal P} \int_{-\infty}^{\infty} \frac{i}{\pi (x-y)}\psi(y) dy
\label{kinetica} \\
&=& -\frac{1}{\pi}
{\cal P} \int_{-\infty}^{\infty} \frac{i}{(x-y)^2}\psi(y) dy\ .
\label{kineticb}
\end{eqnarray}
\label{kinetic}
\end{mathletters}
The symbol ${\cal P}$ stands for the principal value which has
to be taken. If one has to discretize the singular operator
$H_{\rm kin}$, the most reliable approach to follow is to go back a 
few steps using $\omega(k) = v_S|\sin(ka)|/a$ in $k$-space and then 
transform this to real space yielding 
\begin{equation}
\label{kinetic_discrete}
H_{\rm kin} \psi(r_i) = -\frac{1}{\pi} \sum_i \frac{4a\psi(r_i)}{4r_i^2-1}
\end{equation}
where $a$ is the lattice constant.

If we want to treat a chain end, a modification turns out
to be necessary. Calculating with Eq.(\ref{kinetic}) the spinon wave
function for an undimerized odd size chain one finds that the 
spinon is strongly repelled from the borders. This is similar to
the problem of having a quantum mechanical particle in a finite
box where the wave functions are sine-like with nodes at the
wall positions. It is due to the fact that the kinetic energy
cannot be fully satisfied close to a wall since hopping 
processes through the wall are not possible.
However, analyzing the numerical results (see for instance the uniform
susceptibility $\chi^{\rm u}_i$ in Fig.9(b) of
Ref.\cite{lauka98}) one realizes that the probability of finding the
spinon at a given site is approximately constant. We interprete here
the  uniform susceptibility $\chi^{\rm u}_i$ as calculated by
Laukamp {\it et al.}\cite{lauka98} as a measure for the probability to find 
the spinon. From this observation we conclude that the 
walls not only truncate the
hopping processes but that they induce also another effect. To account
for the fact that an approximately constant spinon wave function is 
 the ground state wave function for the problem without dimerization we modify
Eq.(\ref{kinetic}) to 
\begin{equation}
H'_{\rm kin} \psi(x) 
= -\frac{1}{\pi}
{\cal P} \int_{-\infty}^{\infty} \frac{i}{(x-y)^2}\psi(y) dy -
\frac{\psi(x)}{\pi x}\ .
\label{bordercorr}
\end{equation}
The effect of this modification is most easily understood by going back
to Eq.(\ref{kinetica}) and exchanging the sequence of differentiation
and convolution. A constant wave function yields a $\delta$-function
due to the initial jump. The resulting term is compensated by the
$1/x$ term in Eq.(\ref{bordercorr}). We like to stress that the modification is
motivated only phenomenologically. It would be desirable to have a more
microscopic justification as well.

Equivalently, the discrete version of Eq.(\ref{bordercorr}) reads
\begin{equation}
H'_{\rm kin} \psi(r_i) = -\frac{1}{\pi} \sum_j 
\frac{4a\psi(r_j)}{4(r_i-r_j)^2-1} -\frac{2\psi(r_i)}{\pi(2r_i-1)}
\label{bordercorr_discrete}
\end{equation}
if the site counting starts with $i=1$ at the border.
Thus, for chain ends with small frustration we will use
Eq.(\ref{bordercorr}) as kinetic energy. For bulk problems we will use 
Eq.(\ref{kinetic}).  The corresponding discrete versions are
Eq.(\ref{bordercorr_discrete}) and Eq.(\ref{kinetic_discrete}), respectively.

\begin{figure}
  \center{\includegraphics[width=8cm]{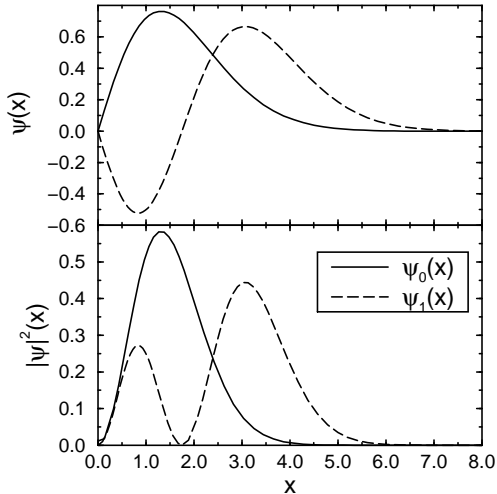}}
\caption[]{Ground state and first excited state wave functions $\psi_0(x)$
and $\psi_1(x)$ using a linear potential $V(x)=x$ and a quadratic kinetic
energy $H_{\rm kin} =k^2$. The corresponding eigen energies are
$e_0=-z_0=2.338$ and $e_1=-z_1=4.088$.
\label{fig-ai}}
\end{figure}

\begin{figure}
 \center{\includegraphics[width=8cm]{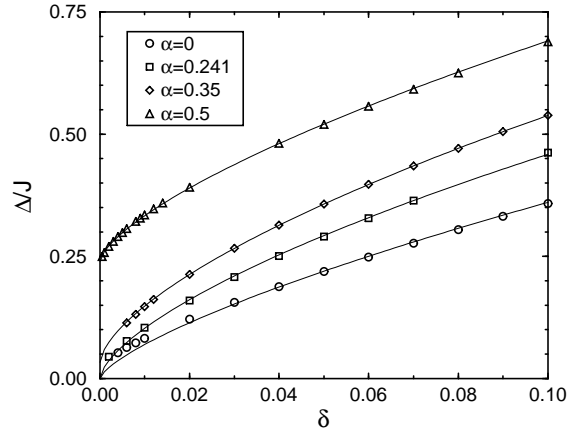}}
\caption[]{Triplet gap as function of the dimerization $\delta$ for
various values of the frustration $\alpha$. Symbols are DMRG results;
the solid lines are simple power law fits 
$\Delta \approx \Delta_0 + \Delta_1 \delta^\nu$
with the parameters ($\alpha$: $\Delta_0$, $\Delta_1$, $\nu$) 
(0: 0, $1.57$, $0.65$); ($0.241$: 0, $2.05$, $0.65$);
($0.35$: $0.033$, $2.216$, $0.642$); ($0.5$: $0.24$, $2.183$, $0.685$).
The absolute error is $10^{-4}J$ at most.
\label{fig-gap}}
\end{figure}

Thus far, we have discussed only the kinetic energy. Considering now the
potential 
energy we first consider regime (i) and in particular the 
Majumdar-Ghosh point $\alpha=0.5$ for which the ground state is
identical to the short-range RVB state. As noted previously,
e.g. \cite{khoms96,affle97}, the confining potential is a linear one 
in this regime.
The potential due to the dimerization $\delta H_{\rm D}$ with
\begin{equation}
\label{hdimer}
H_{\rm D} =  \sum_{j=1}^\infty (-1)^j \vec{S}_{2j-1} \vec{S}_{2j} 
\end{equation}
is given by
\begin{equation}
V(2i+1) = \delta 
\left( \langle i| H _{\rm D}| i\rangle - \langle 1| H _{\rm D}| 1\rangle
\right) \ .
\end{equation}
By inspecting
Fig.\ref{fig-idea}(a) and (c) we find that $V(2i+1)$
increases
by $(3/2) J $ if the spinon hops once, since one singlet
on the strong bonds is lost on the right side whereas one singlet
is inserted on the weak bonds on the left side.
Thus we have 
\begin{equation}
\label{potmg}
V(2i+1) = \delta \frac{3}{2} J i \ .
\end{equation}
This allows us to propose the following Schr\"odinger
equation for the motion of the spinon
\begin{equation}
\label{schrod1}
E \psi(x) = -\frac{J}{2m} \frac{\partial^2}{\partial x^2} \psi(x)
+ \frac{3\delta J}{4}x \psi(x)
\end{equation}
with the restriction $x \ge 0$. The value of $m$ can be found from
a variational investigation of a single spinon and is found to
be approximately $m= (1+7/\sqrt{65})^{-1}\approx 0.535 $ 
\cite{caspe84,mulle98}. Rescaling Eq.(\ref{schrod1}) by $x = \xi y$ with 
\begin{equation}
\label{xi1}
\xi= (3m\delta/2)^{-1/3}
\end{equation}
 (the site spacing is set to unity) yields
\begin{equation}
\label{schrod11}
E \tilde\psi(y) = J\left(\frac{(3\delta/4)^2}{2m}\right)^{1/3} 
\left(-\frac{\partial^2}{\partial y^2} \tilde\psi(y)
+ y \tilde\psi(y) \right)\ .
\end{equation}
The linear differential equation (\ref{schrod1}) with the boundary
condition $\tilde\psi(0)=0$ is solved by shifted Airy functions \cite{abram64}
\begin{mathletters}
\begin{eqnarray}
0 &=& -\frac{\partial^2}{\partial y^2} {\rm Ai}(y)
+ y {\rm Ai}(y) \quad \Rightarrow
\\
- z_i {\rm Ai}(y+z_i) &=& -\frac{\partial^2}{\partial y^2} {\rm Ai}(y+z_i)
+ y {\rm Ai}(y+z_i)
\end{eqnarray}
\end{mathletters}
where the $z_i$ are the zeros of ${\rm Ai}(y)$ which define
by $e_i = - z_i$ the eigenenergies of the rescaled problem.
 The normalization is
given by $\tilde\psi(y) = {\rm Ai}(y+z_i)/|{\rm Ai}'(y+z_i)|$.
Summarizing, using Eq.(\ref{xi1}) we have obtained
\begin{mathletters}
\label{result1}
\begin{eqnarray}
E_i &=& -z_i J\left(\frac{(3\delta/4)^2}{2m}\right)^{1/3} 
\\
\psi(x) &=& \frac{{\rm Ai}(x/\xi+z_i)}{\sqrt{\xi} |{\rm Ai}'(y+z_i)|}\ .
\end{eqnarray}
\end{mathletters}
The form of the wave functions are given in Fig.\ref{fig-ai} where 
$\xi$ is made equal to unity. 

A linear potential together with
a quadratic kinetic energy leads to energies proportional to
$\delta^{2/3}$ if $\delta$ measures the potential strength.
Such a behavior is actually observed in  the dependence of the triplet
gap $\Delta_{\rm trip}(\delta)$ on the dimerization
 $\Delta_{\rm trip}(\delta)-\Delta_{\rm trip}(0) \propto \delta^{2/3}$,
see Fig.\ref{fig-gap} and \cite{yokoy97}. The corresponding 
calculation   is  almost equivalent to the above one since the
triplet state can be viewed as two parallel spinons bound together by
the same potential. Then, the coordinate $x$ corresponds to the
relative coordinate and the mass $m$ has to be replaced by the
relative mass $\mu=m/2$ since one has two kinetic energy
contributions, see e.g. \cite{affle97,els98}.

In the light of the success of a simple quantum mechanical picture
for the case of large frustration we come back to case (ii) of 
subcritical frustration. Before entering the discussion we 
emphasize that without a gap there is {\it a priori} only a less good
justification for a one-particle description since the length scale
of the object carrying the $S=1/2$ is the same as the spatial
extent of the bound wave functions. Yet it is interesting to see
that the simple model works well qualitatively and to a certain
extent even quantitatively.

For subcritical frustration, we have two important pieces of 
information. The kinetic energy is linear in $k$, not quadratic. But
the exponent of the (triplet) gap growth with dimerization
 is also 2/3 or close to it \cite{cross79,chitr95,uhrig96b,yokoy97}
($\Delta_{\rm trip} \propto \delta^{2/3}$). To illustrate 
this point we show in Fig.\ref{fig-gap} gap data for four
different values of the frustration.
Power law fits with exponents
close to 2/3 describe fairly well the gap growth in the 
subcritical and the supercritical frustration regime. The
appropriate fit parameters, however, depend on the fit interval chosen.
It is known that logarithmic corrections make it in general
very difficult to observe the asymptotic behavior. Our estimations  for
the binding length (see below) provide additional information on how small 
the dimerization and how large the system have to be
in order for the asymptotic behavior to be reached.

Thus far, we keep the information that the confining potential cannot be
linear in the subcritical frustration regime.
 Revisiting the scaling that mapped Eq.(\ref{schrod1}) to the
universal form Eq.(\ref{schrod11}) shows that the potential must 
grow as a square root $V(x) \propto \sqrt{x}$ in order to retain
the exponent 2/3. This seems  a very plausible conjecture
which will be corroborated by a direct calculation of the potential.

In order to evaluate the potential energy between the chain end 
and the spinon we use ideas of Talstra et
al. \cite{talst95,talst97}. These authors have shown that a spinon state
can be generated to 98\% overlap by inserting a single spin in a
spin chain which is otherwise in its ground state. If we take this
idea over to our problem of a spinon and a chain end we may assume
that the systems between the chain end and the spinon is in its undimerized
ground state which is the ground state of 
a finite piece of chain with open boundary 
condition. We have to calculate the expectation value of the 
dimerization operator $H_{\rm D}$ (\ref{hdimer}) in this undimerized
ground state in order to obtain an estimate for the potential. 
A small refinement is actually
necessary since we are only interested in the expectation value
of $H_{\rm D}$ with respect to the bulk limit. So we set
$V(2i+1)=\langle H'_{\rm D}\rangle_{2i}$ where the subscript $2i$
refers to the  length of the finite piece of chain and
\begin{equation}
H'_{\rm D} = \sum_{j=1}^\infty (-1)^j (\vec{S}_{2j-1} \vec{S}_{2j} -
\langle\vec{S}_{1} \vec{S}_{2}\rangle_{\rm bulk})\ .
\end{equation}
From this equation one sees that the reference to the bulk limit 
yields only a finite offset. 

Note that calculating $\langle H'_{\rm D}\rangle$
with respect to the undimerized ground state yielding a
potential $V(x) \propto \delta$ is in the spirit of degenerate first
order perturbation theory. The effect of the perturbation, here dimerization,
in the subspace of the unperturbed elementary excitations is considered.
Even though these excitations are not really degenerate
they are arbitrarily close in energy so that none of them can be
discarded.

\begin{figure}
\begin{picture}(8.2,9)
\put(0,0.1){\includegraphics[width=8cm]{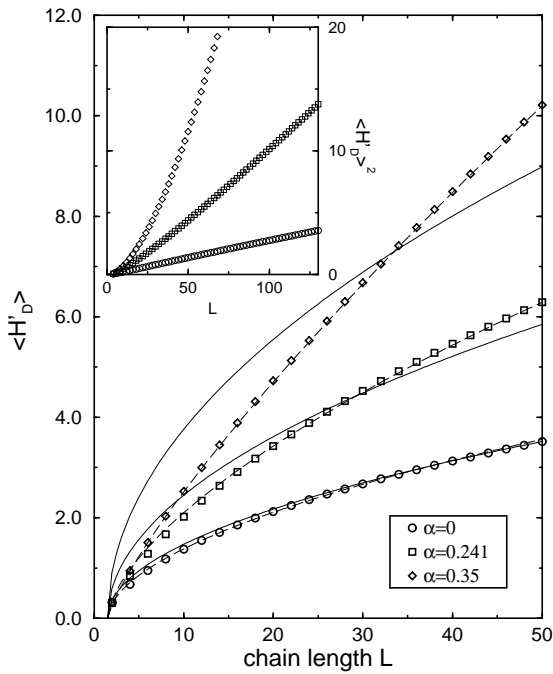}}
\end{picture}
\caption[]{Expectation value of $H'_{\rm D}$ as explained in the main
text for three values of $\alpha$. The solid lines are square root fits
$\propto \sqrt{L-1.5}$ with the prefactors $0.505$, $0.84$, and $1.29$
for $\alpha=0$, $0.241$, and $0.35$.
The offset $1.5$ is chosen for the improvement
of the fits. The dashed lines are power law fits $\propto
(L-1.5)^\beta$ with the exponents and prefactors
$0.544$, $0.431$ for $\alpha=0$, $0.630$, $0.545$ for $\alpha=0.241$, and
 $0.804$, $0.450$ for $\alpha=0.35$. Inset: Square of the same data for longer
chains.
\label{fig-pot}}
\end{figure}

The calculation of $\langle H'_{\rm D}\rangle$ is a perfect task
 for the DMRG approach \cite{white92,white93}, especially since we
deal with open boundary conditions. The results are shown in 
Fig.\ref{fig-pot}. We use the infinite system algorithm \cite{white93},
keeping 100 states in each step. The error in the energies due to basis
truncation is found to be smaller than $10^{-5}$.
One clearly observes a sublinear increase of $\langle H'_{\rm D}\rangle$
with the chain length. Without frustration ($\alpha=0$) a square root
power law fits the data perfectly if a small offset on the $x$-axis is
taken into account. For larger frustration larger exponents yield better fits.
The inset, however, shows that for longer chain lengths ($L>50$), the square 
root behavior is recovered for $\alpha\le \alpha_c$
which can be seen from the linear behavior of the squared values.
 The frustration value $\alpha=0.35$ is included to show that for this
value relatively close to the critical one a linear behavior 
$\langle H'_{\rm D}\rangle \propto L$ cannot yet be seen.

Another evidence that for $\alpha<\alpha_c$ the expectation value
$\langle H'_{\rm D}\rangle$ rises proportional to $\sqrt{L}$
stems from the dimension 1/2 of the dimerization operator \cite{cross79}.
By integration $\int^L dx/\sqrt{x} \propto \sqrt{L}$ follows the 
conjectured behaviour.

\begin{figure}
 \center{\includegraphics[width=8cm]{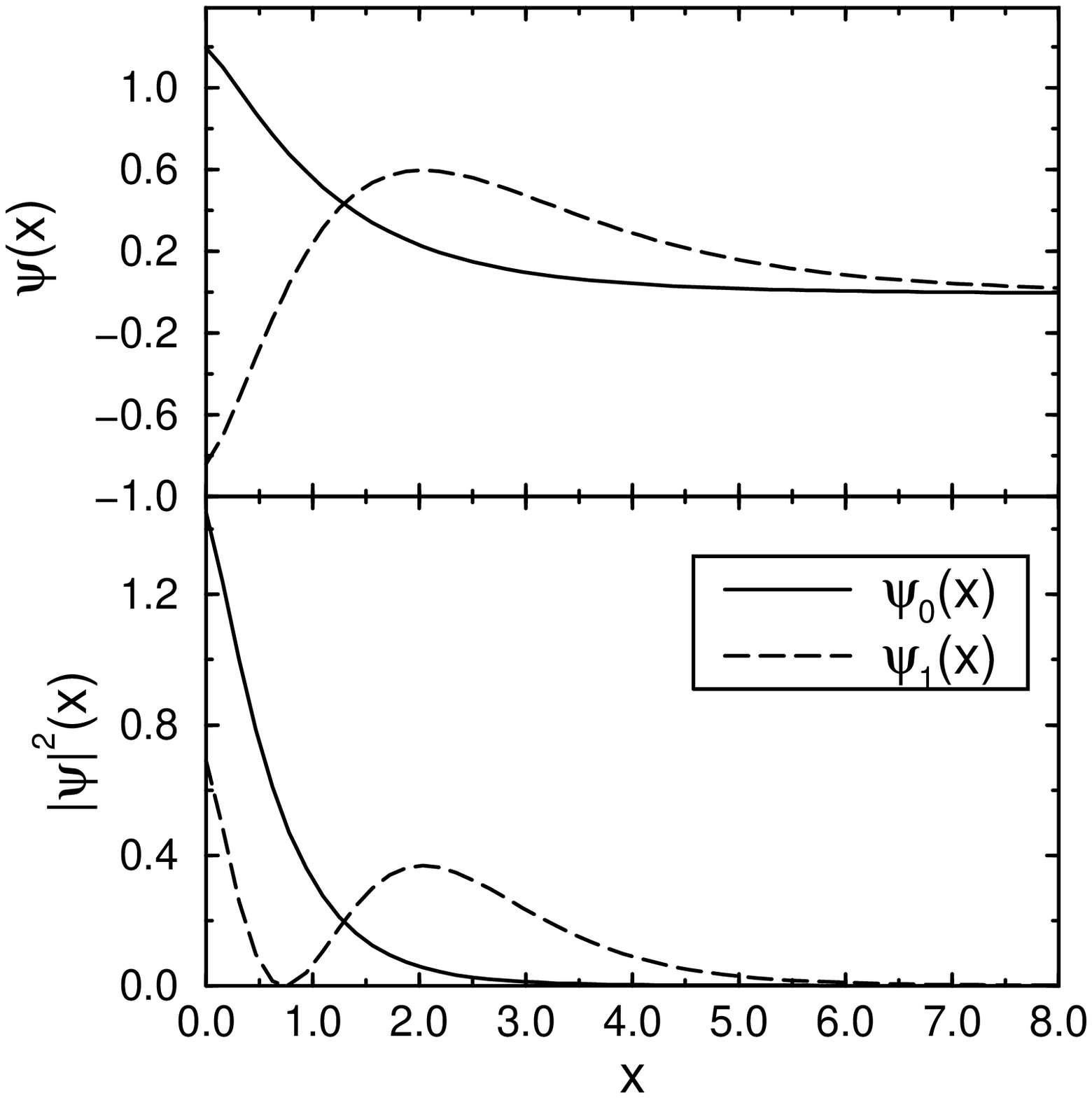}}
\caption[]{Ground state and first excited state wave functions $\psi_0(x)$
and $\psi_1(x)$ using a square root potential $V(x)=\sqrt{x}$ 
and linear kinetic
energy $H_{\rm kin} = |k|$ plus border corrections (see Eq.(\ref{bordercorr})).
 The corresponding energies are $e_0=1.049$ and $e_1=2.040$.
\label{fig-wfos}}
\end{figure}

So our conjecture of a square root confining potential for lower
frustrations is corroborated by direct calculations. We are now in
the position to perform an analysis as in Eqs.(\ref{schrod1} -\ref{result1}).
Starting from the generalized Schr\"odinger equation with
the border effect corrected kinetic part $H'_{\rm kin}$ Eq.(\ref{bordercorr})
\begin{equation}
\label{schrod2}
E\psi = v_{\rm S} H_{\rm kin} [\psi] + \delta A \sqrt{x} \psi
\end{equation}
we rescale by $x=\xi y$ with
\begin{equation}
\label{xi2}
\xi = \left(\frac{v_{\rm S}}{\delta A}\right)^{2/3}
\end{equation}
where $v_{\rm S}$ is the spin wave velocity and $A$ some constant
which can be deduced from fits to data as in Fig.\ref{fig-pot}.
Eq.(\ref{schrod2}) becomes
\begin{equation}
\label{schrod21}
e_i \psi_i = H'_{\rm kin} [\psi_i] + \sqrt{y} \psi_i
\end{equation}
from which the energies are found using
\begin{equation}
\label{schrod22}
E_i = e_i (\delta^2 A^2 v_{\rm S})^{1/3}\ .
\end{equation}
The resulting two first wave functions are shown in Fig.\ref{fig-wfos}.
The dimensionless energies are $e_0 = 1.049$ and $e_1 = 2.040$.
Equation (\ref{schrod21}) is solved numerically by equidistant 
discretization and fast Fourier transform with up to $2^{19}$ points.
Due to the singular behavior of Eq.(\ref{schrod21}) the precision is only
of about $10^{-3}$. But this is sufficient for the intuitive description
 that is being proposed  here.

Note that due to the border correction Eq.(\ref{bordercorr})
the wave functions do not vanish at the chain end, i.e. at $x=0$.
Furthermore, the wave functions decay rapidly as $x\to \infty$,
but not following an exponential form, e.g. \cite{fukuy96,mikes97},
 but rather a power law 
 $\psi(x) \propto 1/x^3$. This is due to the highly non-local 
character of the kinetic
energy Eq.(\ref{kinetic}). Thereby, we do not want to claim that the
decay for large $x$ follows indeed a power law. The picture we are
presenting does not contain the generation of further 
spinon-antispinon pairs since it is a quantum mechanical, {\em single}
particle scenario. It is the pair creation possible for energies larger
than the gap $\Delta_{\rm trip}$ which eventually leads to
 the exponential decay of the spinon probability. 
Our picture works best at intermediate
distances as we will show in comparison to the DMRG data in the next section.

\section{Results: Chain Ends}

Now we are in the position to compare results of bound spinon
calculations directly against numerical results.
\begin{figure}[htb]
\begin{picture}(8.2,12.2)
 \put(-2.5,0){\includegraphics[width=14cm]{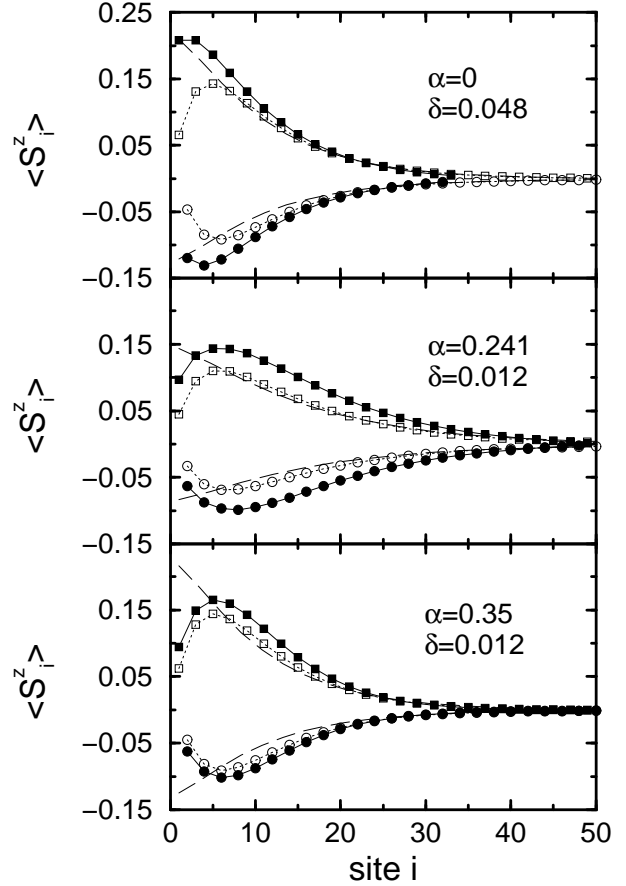}}
\end{picture}
\caption[]{Local magnetizations at dimerized chain ends with weak bonds for
various values of $\alpha$ and $\delta$.
 $S^z_{\rm tot}$ is set to $1/2$. Squares:
odd sites; circles: even sites; filled symbol: DMRG; open symbols:
discrete bound spinon calculation; dashed lines continuous bound spinon
calculation.
\label{fig-vgl}}
\end{figure} 
 The latter are obtained using the   DMRG technique \cite{white92,white93} 
for open boundary conditions with between $m=24$ to 32 states kept in 
the iterations. The results are shown in
Fig.\ref{fig-vgl}. Quantum mechanical bound spinon calculations
are carried out in two ways. One is the continuum one relying on
Eqs.(\ref{resodd1},\ref{reseven1},\ref{bordercorr}),
the input for the spin wave velocity
Eq.(\ref{spinwave}) and the potential parameters for a square root
potential (see caption of Fig.\ref{fig-pot}). This means that one 
takes the result for $|\psi(x)|^2$ in Fig.\ref{fig-wfos} and rescales 
$|\psi(x)|^2 \to|\psi(x/\xi)|^2/\xi  $ with $\xi$ from
Eq.(\ref{xi2}). The relevant values are
$\xi=16.1$, $23.5$, and $15.6$ for the uppermost to the lowermost
panel in Fig.\ref{fig-vgl}. The results are shown as dashed curves. 

The other one is a discrete treatment relying on
Eqs. (\ref{resodd2},\ref{reseven2},\ref{bordercorr_discrete}).
The results are depicted with open symbols.
The values of the potential are those read off Fig.\ref{fig-pot}
with $V(2i+1)=\langle H'_{\rm D}(L=2i) \rangle$. For simplicity 
the very good power law fits (dashed lines in Fig.\ref{fig-pot})
are used for the numerics.

The agreement between the DMRG results and the bound spinon
model is very good, especially in view of the simplicity of the model.
The shape of the curves are very well described by the
bound spinon results. In particular, the correct binding length
is predicted. This is in particular interesting since we know that the
quantum mechanical model does not display an exponential
decay. This means that the exponential tail matters only at larger
lengths. For intermediate lengths, where the main weight is found,
the bound spinon calculations work very well. The main feature that
is included by the disrete calculation is the decrease of the amplitude
for the first four to five sites. It results from the properties
of Eqs.(\ref{resodd2},\ref{reseven2}) and can thus be attributed to
border effects which were neglected in Eqs.(\ref{resodd1},\ref{reseven1})
but properly taken into account in Eqs.(\ref{resodd2},\ref{reseven2}).
 It is {\em not} due to the dynamics of the problem.
The very good agreement of the open-symbol curves and the long-dashed
curves for larger distances underlines the applicability of the approximation
(\ref{norm}).

The overall amplitude is larger than the one predicted by the
bound spinon model. Since the basis we used included only the 
shortest range singlets it is not surprising that the true
antiferromagnetic correlations are in fact larger. Thus we view the
bound spinon results in this respect as a lower bound for the
antiferromagnetic correlations. But it is interesting that even
for the amplitudes the agreement is reasonable for correlation lengths $\xi$
around 15. For larger values of $\xi$ (middle panel in Fig.\ref{fig-vgl})
the agreement becomes not as good.

Another point that can be addressed easily within the bound spinon
model are excitation energies, namely the energy difference $E_1-E_0$
(\ref{schrod22})
between the ground state $\psi_0$ and the first excitetd state
$\psi_1$, see Fig.\ref{fig-wfos}. The continuum (discrete) calculation yields
$0.97$ ($0.10$), $0.048$ ($0.055$), and $0.061$ ($0.073$) for the three
cases in Fig.\ref{fig-vgl} in descending order. It is not surprising
that the continuum calculation provides lower values since it assumes
a  potential with smaller exponent, see Fig.\ref{fig-pot}), 
so that the distances between the energy levels become smaller.
A smaller exponent implies that the potential increase for smaller
arguments is larger than for larger arguments. Thus the smaller of
two consecutive eigenenergies is lifted with respect to the larger one
since the wave function belonging to the smaller eigenenergy is more
localized. This effect  explains also why the
deviation is the largest for $\alpha=0.35$ where the deviation of the optimum
fit exponent to $1/2$ is largest.

It was mentioned before that the occurrence of the triplet gap
in dimerized systems can be viewed also as a binding phenomenon, see e.g.
\cite{affle97,els98}.
In this case, two spinons bind. However, since only the relative
coordinate matters, this problem is almost equivalent
 to the binding of a single spinon to a chain end.
The main difference is that the kinetic energy is doubled since both
interaction partners move.
 So it is understandable that the ratio
$(E_1-E_0)/\Delta_{\rm trip}$ is fairly constant.
The triplet gaps $\Delta_{\rm trip}$ as read off Fig.\ref{fig-gap} are
$0.225$, $0.116$, and $0.162$. The ratios  of the continuum (discrete)
values  are  $0.43$ ($0.46$),$0.42$ ($0.47$), and $0.37$ ($0.45$).
Similar to the result found previously ($0.6$) \cite{els98} the transition from
the ground state to the first excited one is at about half the
triplet gap. This implies that it should be observable in scattering
experiments as a rather sharp feature within the gap. Indeed, Raman
scattering results have revealed such a feature in Zn-doped  CuGeO$_3$ 
\cite{els98} even though the energy is higher than the one-dimensional
model predicts. 

It is clear that at energies $E_0+\Delta_{\rm trip}$ pair production
becomes possible as it is known in QCD. The gedanken experiment of separating
quarks in space in spite of the confinement
 results in the production of new quark pairs.
In our situation any excited bound state can decay by
producing a pair of spinons at low values of momentum
if its energy fulfills $E_i \ge E_0+\Delta_{\rm trip}$.
This mechanism gives rise to a continuum which sets in at
$E_0+\Delta_{\rm trip}$. Thus higher energy bound states are no longer
true eigenstates of the problem but have a finite lifetime 
since they might decay.
The discrete calculation actually shows that $E_4$ and higher are no 
longer stable. This means that besides $E_1$, also $E_2$ and $E_3$
should exist as distinct sharp modes. Experimentally there is so far
no indication for these higher modes. But it should be borne in mind that
any deviation from pure $d=1$ behavior tends to decrease the binding energies.
Bound states are shifted closer to the continuum or even disappear
in the continuum due to higher dimensional effects, see also
 \cite{els97,els98}.

\begin{figure}
 \center{\includegraphics[width=8cm]{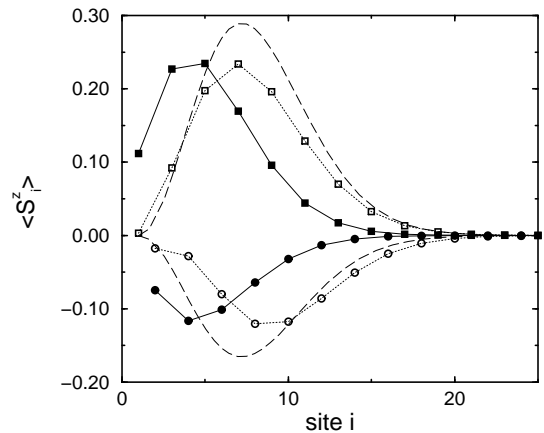}}
\caption[]{Same as  Fig.\ref{fig-vgl} 
but for $\alpha=0.5$ and $\delta=0.012$.
\label{fig-mg}}
\end{figure} 
In Fig.\ref{fig-mg} we show the same results as in Fig.\ref{fig-vgl}
but for the Majumdar-Ghosh point $\alpha=0.5$. They are based on
Eqs.(\ref{schrod1},\ref{resodd1},\ref{reseven1}) in the continuum case.
In the discrete case we used Eqs.(\ref{potmg},\ref{resodd2},\ref{reseven2}).
The discrete kinetic energy used is nearest neighbor hopping such that
the quadratic minimum is the same as for Eq.(\ref{schrod1}).

One should expect 
an improved agreement since the kind of basis we are working with
is particularly suited for the Majumdar-Ghosh point \cite{shast81,mulle98}.
The agreement, however, is not very good. It becomes decisively better
 if the bound spinon results are shifted by about three lattice
spacing. Probably dynamic border effects not taken
into consideration by the basic assumptions concerning the kinetic and
the potential energy (focused on the low energy behavior) are important.
The binding length $\xi$ for
the Majumdar-Ghosh model according to Eq.(\ref{xi1}) is only $4.7$.
Thus a low-energy approach might be insufficient for such a small
binding length. 

Interestingly, Eggert and Affleck found  in a comparison
of continuum field theoretical results with numerical data at $\delta=0$ 
also that
the agreement is much better if the continuum results are shifted
by two sites towards the chain end \cite{egger95}. 
An understanding of either of these
observations would help to understand the other.

\section{Results: Kink Defects}

Spin-Peierls systems are characterized by a coupling of magnetic
(quasi) one-dimensional spin degrees of freedom to the lattice degrees of
 freedom, i.e.\ vibrating
distortions (for reviews see \cite{bray83,bouch96}).
 At low temperature
 and zero  magnetic field the system is in the gapped dimerized (D) phase where
 the coupling strength $J_i$ of the spin chain alternates from site to site. 
As the magnetic field is increased the gap becomes smaller and at a critical
field $H_c$ the phase changes to an incommensurably modulated (I) phase.
This I phase can be viewed as a soliton lattice where equally spaced
$\tanh$-like zeros of the distortions $\delta_i$ occur. In the vicinity
of each zero of the distortion a spinon $S=1/2$ is localized, see 
\cite{schon98} and references therein.

\begin{figure}
\begin{picture}(8.2,6.8)
\put(0,0){\includegraphics[width=8cm]{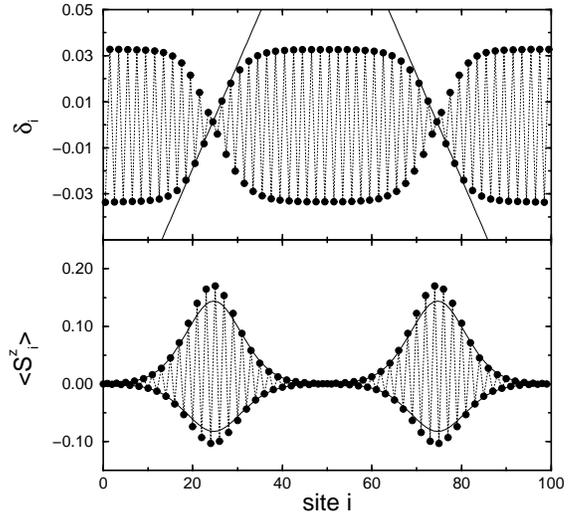}}
\end{picture}
\caption[]{Two solitons in a hundred site chain, i.e. $m=1/100$,
 with periodic boundary
conditions are studied (parameters $\alpha=0.35$, $K=8.7J$).
In the upper panel, results from the self-consistent determined 
distortions $\delta_i$ (circles)
which minimize $\langle H\rangle$ in (\ref{Hamilton}) are shown.
They are depicted between the sites since they belong to bonds.
The straight lines are fits to the linear rising distortions
with slope $s=0.0045$. The lower panel displays the resulting
local magnetizations (circles, DMRG calculation) and the bound spinon
result (solid lines). For details see main text. 
\label{fig-kink}}
\end{figure}

In Fig.\ref{fig-kink} the soliton lattice is illustrated. These results are
found from an adiabatic calculation where the distortions are treated
statically. Most of the investigations of the I 
phase assume static distortions. 
The Hamiltonian $H$ from which Fig.\ref{fig-kink}
is computed by minimization of the ground state energy with respect
to $\{ \delta_i \}$ reads \cite{schon98}
\begin{mathletters}
\label{Hamilton}
\begin{eqnarray}
 H & = &  H_{\rm chain} +  H_{\rm Zeeman} + 
E_{\rm  elast}
\\ 
 H_{\rm chain} & = & \sum_{i=1}^{L} \left( J_i
  {\vec S}_i \cdot {\vec S}_{i+1} +J \alpha \, {\vec S}_i \cdot {\vec S}_{i+2}
\right) 
\\ 
 H_{\rm Zeeman} & = & g\mu_{\rm B} H S_z \ ,
\\
E_{\rm elast} & = & \frac{K}{2} \sum_i \delta_i^2  \ , 
\\ 
J_i & =& J(1+\delta_i) \ ,
\end{eqnarray}
\end{mathletters}
where $\alpha$ denotes the relative frustration and $S_z$ is the $z$
component of the total spin of the $L$-site chain. The last two terms
in Eq.(\ref{Hamilton}) are the Zeeman energy and the elastic energy 
associated to the lattice distortion. A site independent spring constant
$K$ is used.

There are regions around site, e.g., 50 where the distortion alternates as
in the {D} phase. The corresponding local magnetization is essentially
zero. Close to the linear zeros of the distortions at about sites 25 and 75
one observes a strongly alternating local magnetization which adds up to
1/2: $\sum_{i=0}^{50}\langle S^z_i\rangle = 1/2$. The distribution of
local magnetizations is observable by NMR \cite{fagot96} and gives interesting
information on the {I} phase and on the validity of the adiabatic
treatment \cite{uhrig98a}. An interesting observation
is that the ratio between the parallel magnetizations to the antiparallel
magnetization at (roughly) the same sites is 7:4 as implied by 
Eqs.(\ref{resodd1},\ref{reseven1}).

In this paper we are not interested in the self-consistent minimization of
(\ref{Hamilton}) but we take the  $\{\delta_i\}$ as given and  use
the quantum mechanical picture developed above to show that the localized
alternating magnetization in the vicinity of a zero of the distortion
results from a bound spinon. The zero of the distortion is viewed
as a kink defect. Since the $\delta_i$ on the odd (even) bonds change
sign at the zero there is no simple short-range singlet pattern which
ensures singlets at the stronger bonds without any free spin. Insofar
the situation is comparable to the one at chain ends with weak bonds at
the ends. The spinon bound to a kink defect is even simpler than the
spinon bound to a chain end since no particular boundary effects on the
kinetic energy need to be taken into account. Thus we use the continuum
description based on Eqs.(\ref{resodd1},\ref{reseven1},\ref{kinetic}).

The derivation of the attractive potential requires an additional assumption
in order to treat also site dependent distortions. We find it natural to
assume that the chain end potential $V(x)$ as determined from Fig.\ref{fig-pot}
results from expectation values 
$\langle \vec{S}_i\vec{S}_{i+1} \rangle - 
\langle \vec{S}_i\vec{S}_{i+1} \rangle_{\rm bulk}$ which decay like a power
law with increasing distance from the ends
\begin{eqnarray}
V(x) & = & \delta a x^b \nonumber \\
\label{general-pot1}
 & = & \frac{ab}{2} \int_0^x (\delta y^{b-1} + \delta (x-y)^{b-1})dy\ ,
\end{eqnarray}
where we took into account that the chain end {\it and} the position
of spinon constitute borders. The effect of these borders is treated
approximatively as additive, i.e. as independent. 
 A general power law with exponent $b$ and prefactor $a$ is dealt with.
 The advantage of 
Eq.(\ref{general-pot1}) is that the distortion can be made site dependent
$\delta \to \delta(x)$. In the vicinity of the kink defect, $\delta(x) = s x$
is a good description where $s$ is an appropriately determined slope, see
upper panel in Fig.\ref{fig-kink}. A short calculation then leads to
\begin{eqnarray}\nonumber
V(x)_{\rm kink} &=& \frac{abs}{2} \int_0^{|x|} (y y^{b-1} + 
y (|x|-y)^{b-1})dy\\
&=& \frac{as}{2} |x|^{b+1}\ .
\label{pot-kink}
\end{eqnarray}
\begin{figure}
 \center{\includegraphics[width=8cm]{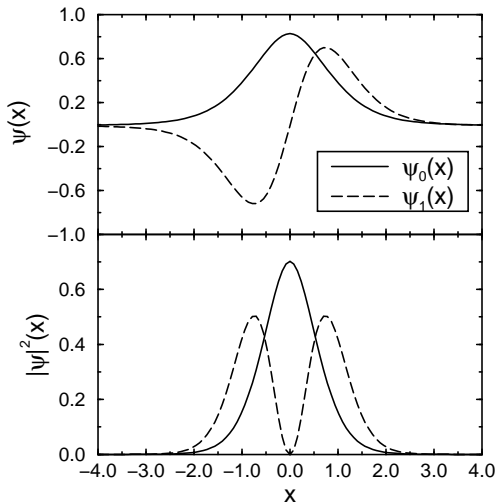}}
\caption[]{Ground state and first excited state wave functions $\psi_0(x)$
and $\psi_1(x)$ using the integrated square root potential
 $V(x)= |x|^{3/2}$ and the linear kinetic
energy $H_{\rm kin} = |k|$ (see Eq.(\ref{kinetic})).
 The corresponding energies are $e_0=1.051$ and $e_1=2.310$.
\label{fig-wfds}}
\end{figure} 
Thus we obtain again a power law confining potential, the exponent of 
which is increased by one compared with the site-independent distortion.
Moreover, the potential does not depend on whether the spinon moves to
the right or to the left of the zero.

In Fig.\ref{fig-wfds} a generic result for the localized state at a kink
defect is shown. All parameters are set to unity. But rescaling permits,
as before, to obtain the general relations
\begin{mathletters}
\begin{eqnarray}
\label{xi3}
\xi &= & \left( \frac{2v_S}{as}\right)^{1/(2+b)} \\
E_i &= & e_i \left(v_S^{1+b} \frac{as}{2}\right)^{1/(2+b)}
\end{eqnarray}
\end{mathletters}
The solid lines in the lower panel in Fig.\ref{fig-kink} depict
the convincingly good agreement between the bound spinon picture
and the DMRG calculation. The value $s=0.0045$ and the parameters of the 
dashed line fit to the $\alpha=0.35$ data in Fig.\ref{fig-pot} is used 
for the binding length $\xi$ in (\ref{xi3}).

\section{Summary}

In this work we have developed a quantum mechanical picture 
in the spirit of first order degenerate perturbation theory
which describes the confinement of spinons due to dimerization.
Without dimerization there is no confinement independently of the
value of the frustration $\alpha$.
The picture works below and above the critical frustration of the 
undimerized chain $\alpha_c = 0.241$, even though the justification
below $\alpha_c$ is less good.
The main ingredients of the calculation
 are the known facts concerning the kinetic energy
of the spinons which is linear below $\alpha_c$ and quadratic (plus a mass
term) above $\alpha_c$. From the dimer expectation values close to
the borders of finite chains we deduced a potential $V(x)$ which is 
proportional to the dimerization $\delta$. It was shown that this
potential can be nicely fitted for intermediate distances ($L\approx 50$)
by sublinear power laws  below  and just above $\alpha_c = 0.241$.
For larger distances ($L\to \infty$) our results indicate that $V(x)$
increases like a square root for $\alpha \le \alpha_c$ and linearly
for $\alpha > \alpha_c$. To our knowledge, the fact
 that the confining potential  below $\alpha_c = 0.241$ has to be sublinear
 has not been reported before in the literature.

The quantum mechanical picture describes very well
 the local magnetizations close
to chain ends with weak bonds  for intermediate $\xi$ ($\approx 20$).
A very good agreement is also found for the spinon bound to a kink defect.
The sum of the moduli of the local magnetizations is (at least) enhanced by
a factor 11/3 over the value for a single spin.
For low $\xi$ values the continuum description becomes less reliable.
For  large values of $\xi$ the form of the site dependence of the local 
magnetization is still captured by our quantum mechanical picture.
In particular, binding lengths can be easily estimated using
Eqs.(\ref{xi1},\ref{xi2},\ref{xi3}). The
overall amplitude of the staggered component, however, is in fact larger
for $\delta\to 0$ at $T=0$ \cite{egger95} .

Besides local magnetizations the eigenenergies of the bound states
can be estimated easily within our approach. We find that in truly
one-dimensional systems transitions to the first three excited
states at chain ends should be possible. Any degree of higher dimensionality,
however, will reduce this number.

The formation of gapful triplets can equally be viewed as binding of
two spinons \cite{affle97}. The essential differences to the case of
chain ends are that the site variable denotes the relative coordinate between
the spinons and that the kinetic energy is doubled since both
interaction partners move.

The quantum mechanical picture is scale invariant. This means that for
power law potentials the wave functions and energies can all be scaled 
to one universal case, see Figs.\ref{fig-ai}, \ref{fig-wfos},  and
\ref{fig-wfds}. Thus the quantum mechanical description provides a 
pedestrian approach to perturbations with anomalous dimensions and
exponents. The anomalous exponents result naturally from the scaling behavior
of the potential and the kinetic energy under changes of the length scale.
A dimensional analysis implies already that for $V(r) = a r^b$
and $H_{\rm kin} = c |k|^d$ the characteristic lengths scale like
$\xi \propto (c/a)^{1/(b+d)}$ and the characteristic excitation energies like
$E \propto (a^d c^b)^{1/(b+d)}$.

We like to point out that the basis states of our approach can be refined to
take longer range antiferromagnetic correlations better into account.
To this end longer range singlet pairs must be included and
superposed. Dispersions
and potentials can then be found by concepts similar to those proposed
by Sutherland \cite{suthe88} for short range singlets in $d=2$.

\begin{figure}
\begin{picture}(8.2,1.5)
\put(0,0.3){\includegraphics[width=8cm]{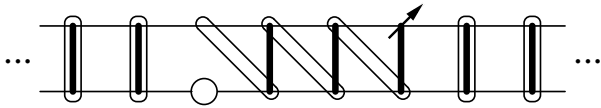}}
\end{picture}
\caption[]{
Sketch of the singlet distribution on a doped ladder with strong bonds 
(thick solid lines) on the rungs. The open eyelets represent singlets.
The circle indicates the defect (nonmagnetic dopant), the arrow the 
generated spinon after three hops. The misaligned diagonal singlets generate
the confining potential.}
\label{fig-ladder}
\end{figure} 
As an outlook to higher dimensions
we present in Fig.\ref{fig-ladder} the leading sketch
of the singlet distribution on a ladder with strong bonds on the
rungs. Ladders are objects of ample investigations. The alternating 
magnetizations \cite{motom96,fukuy96,mikes97} and enhanced antiferromagnetic
correlations \cite{sigri96,iino96} indicate the similarity to open chains.

In Fig.\ref{fig-ladder} the situation is depicted
of the spinon (arrow) generated by
the insertion of the defect (circle) after three hops on the
upper leg. One clearly sees that the motion of such a spinon
is also confined by a monotonic increasing potential. An increasing 
distance of the spinon from its original rung induces an increasing
number of misplaced singlets. This leads to an increase in energy.
From Fig.\ref{fig-ladder} it is clear that the basic concepts of 
spinon motion and confining potential can be extended also to 
other RVB-type spin systems. The results discussed in this paper
are not restricted to chains only.

\section*{Acknowledgment}
We  thank G. Els, P.H.M. van Loosdrecht, G. Martins, and
E. M\"uller-Hartmann for helpful discussions.
One of us (GSU) acknowledges the kind hospitality of the NHMFL, 
Tallahassee,  where a large part of this work was carried out.
We acknowledge financial support by the DFG through the SFB 341
and by the NSF grant DMR-9520776.


\end{document}